\documentclass[12pt]{article}
\usepackage{amsmath}
\usepackage{graphicx,psfrag,epsf}
\usepackage{enumerate}
\usepackage[square,sort,comma,numbers]{natbib}
\usepackage{url} 
\usepackage{EAH}
\usepackage{pbox}

\makeatletter
\newcommand{\longdash}[1][2em]{%
  \makebox[#1]{$\m@th\smash-\mkern-7mu\cleaders\hbox{$\mkern-2mu\smash-\mkern-2mu$}\hfill\mkern-7mu\smash-$}}
\makeatother
\newcommand{\omitskip}{\kern-\arraycolsep}

\newcommand*{\vertbar}{\rule[-1ex]{0.5pt}{2.5ex}}

\newcommand{\blind}{0}

\begin{document}

\def\spacingset#1{\renewcommand{\baselinestretch}%
{#1}\small\normalsize} \spacingset{1}

\if0\blind {
  \title{Comparative Study of Clustering Techniques for Real-Time Dynamic Model Reduction
    \thanks{This material is based upon work supported by the U.S. Department of Energy,
    Office of Science, Office of Advanced Scientific Computing Research, Applied
    Mathematics program.}}
  \author{Emilie Purvine$^\dagger$, Eduardo Cotilla-Sanchez$^\ddagger$,
Mahantesh Halappanavar$^\dagger$,\\
Zhenyu Huang$^\dagger$, Guang Lin$^\circ$, Shuai Lu$^\bullet$,
Shaobu Wang$^\dagger$\\
\small{$^\dagger$Pacific Northwest National Laboratory; \texttt{first-name.last-name@pnnl.gov}}\\
\small{$^\ddagger$Oregon State University; \texttt{ecs@eecs.oregonstate.edu}}\\
\small{$^\circ$Purdue University; \texttt{guanglin@purdue.edu}} \\
\small{$^\bullet$EnerMod; \texttt{shuai.lu@EnerMod.com}} }
  \maketitle
} \fi

\if1\blind {
  \bigskip
  \bigskip
  \bigskip
  \begin{center}
    {\LARGE\bf Comparative Study of Clustering Techniques for Real-Time Dynamic Model Reduction}
  \end{center}
  \medskip
} \fi

\bigskip
\begin{abstract}
Dynamic model reduction in power systems is necessary for improving
computational efficiency. Traditional model reduction using linearized
models or offline analysis is not adequate to capture dynamic behaviors of
the power system, especially with the new mix of intermittent generation
and intelligent consumption making the power system more dynamic and
non-linear. Real-time dynamic model reduction has emerged to fill this
important need. This paper explores using clustering techniques to analyze
real-time phasor measurements to identify groups of generators with similar
behavior, as well as a representative generator from each group for dynamic
model reduction. Two clustering techniques -- graph clustering and
$k$-means -- are considered. These techniques are compared with a
previously developed dynamic model reduction approach using Singular Value
Decomposition. Two sample power grid data sets are used to test these
different model reduction techniques. Based on the algorithms' relative
performance, recommendations are provided for practical use.

  \goaway{Model reduction in the power grid is a necessary task due to the
  heterogeneity and nature of the interactions within the system.
  Representing a group of similarly behaving generators with a single
  \emph{characteristic generator} is one way to reduce a large system to a
  more manageable one. In this paper we compare a variety of model
  reduction techniques which use different types of data to accomplish
  their task. We compare the results to each other and also to the full
  system.}

\end{abstract}


\noindent%
\vfill \hfill {\tiny technometrics tex template (do not remove)}

\newpage
\spacingset{1.45} 
\section{Introduction}
Power engineers rely on simulation to accomplish operation and planning
tasks. Dynamic simulation tools are used extensively to simulate the behavior
of power systems over time subject to disturbances, such as faults, sudden
loss of transmission paths, and loss of generation or load. Good models and
accurate parameters are essential for credible computer simulation. In some
cases, inaccurate computer models and simulations result in optimistic
decisions, which can put the electric infrastructure in jeopardy. The extreme
consequences of such optimistic decisions can be massive outages, such as the
August 1996 western U.S. system breakup \cite{venkat_analysis_2004}. In many
other cases, inaccurate models and simulations lead to pessimistic decisions,
resulting in reduced asset utilization.

Given the power grid's complexity and large footprint, a power company needs
a reduced model for the region outside of its own service territory. The goal
of such model reduction is to reasonably represent the external system with a
simplified smaller model so analysis can be performed more efficiently
\cite{Antoulas01asurvey,wang_dynamic-feature_2014}. This is of particular
interest for real-time power system operation, such as online dynamic
security assessment \cite{sun_online_2007, tiako_real-time_2012,
xue_new_1998}.

Traditionally, model reduction is performed in the steady-state context,
largely ignoring dynamics, or performed offline where the scenarios may be
different from real-time conditions. This has served the power system
reasonably well when the behaviors are more predictable. However, this
practice is no longer adequate as the power system is becoming increasingly
dynamic and non-linear due to the new mix of intermittent generation and
intelligent consumption. Model reduction must evolve to handle these
non-linear and dynamic behaviors.

Dynamic model reduction has been studied extensively. \emph{Coherency} is the
most common concept adopted to identify groups of dynamic devices, e.g.,
generators, for model reduction purposes. Specifically, a group of
generators, $G = \{g_1, \ldots, g_m\}$, is \emph{coherent} if their
difference in voltage angles is constant over time, i.e., there exists a $c
\in \R$ such that $\delta_{g_i}(t) - \delta_{g_j}(t)=c$ for all $t \geq 0$
and $g_i, g_j \in G$. Coherency can be determined using a linearized model
around an operating point \cite{rogers_power_2000, sastry_coherency_1981} or
by analyzing offline simulated dynamics \cite{rudnick_power-system_1981}.
Either way, it is not capable of capturing real-time operating conditions,
which renders the reduced model useless for timely analysis.

Coherency is inherently a clustering-based reduced order model (ROM) method
--- as are the majority of the methods discussed in this paper. However, it is
important to point out that there are many other types of ROM methods.
Subspace projection methods, which find a simpler subspace of the full model
with certain approximation guarantees, are very common. Proper Orthogonal
Decompositions (PODs) \cite{Pinnau2008} and Krylov subspace methods
\cite{Bai20029} are two examples of subspace projections. One method
discussed in this paper, Singular Value Decomposition (SVD), is also an
example of a subspace method. Beyond clustering and subspace methods, there
are also linearization techniques that are used to reduce variable space of
the dynamic power equations directly \cite{Freund1999}.

Recent developments and deployment of phasor technologies present an
opportunity to perform dynamic model reduction in real time because
high-speed phasor measurements capture the majority of the power system
dynamics necessary for power system operation and planning purposes. Ref.
\cite{wang_measurement-based_2012} proposes an SVD-based method that can be
used for real-time dynamic model reduction. This method also preserves a
certain level of non-linearity in the reduced model. Following this line of
research, our team continues developing real-time dynamic model reduction
techniques using graph clustering methods and compares the accuracy of
reduced models with the SVD-based approach, as well as traditional $k$-means
clustering.

\begin{figure}
  \begin{center}
    \includegraphics[width=0.85\linewidth]{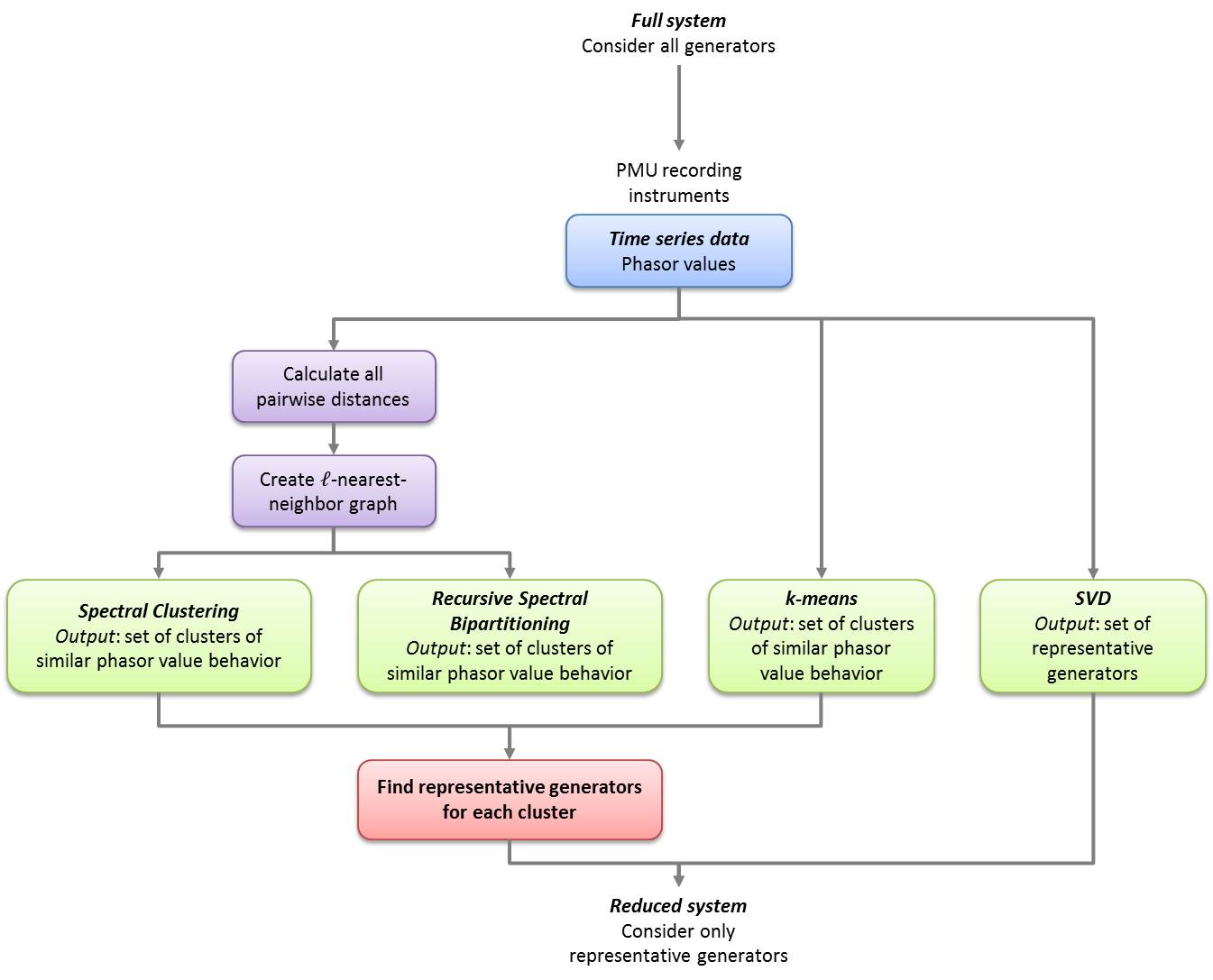}
  \end{center}
  \caption{Workflow showing the four basic reduction methods compared in this paper.}\label{fig:workflow}
\end{figure}

Real-time phasor measurements are used to cluster generators based on similar
behaviors, and representative generators are chosen from each cluster. The
final reduced model contains only the representative generators. We
investigate two graph clustering methods -- recursive spectral bipartitioning
\cite{schaeffer_graph_2007} and spectral clustering
\cite{luxburg_tutorial_2007}. Variants of implementations of these two
methods are tested alongside the SVD and $k$-means algorithms on fault
scenarios within the IEEE 50-machine \cite{canizares_linear_2004} and
16-machine systems. For reference, Figure \ref{fig:workflow} contains a
workflow diagram showing each method found in the remainder of this paper.
Error measurements quantify various levels of accuracy for each method, but
many possess accuracy that is adequate for power system operation and
planning purposes. From a comparative perspective, this paper provides a good
reference point for practical implementations.

The remainder of the paper is organized as follows: Section \ref{sec:data}
describes the system and data used in implementing and comparing the proposed
model reduction methods. Sections \ref{sec:methods} and \ref{sec:reduce}
provide an overview of the clustering techniques and representative generator
identification, including their computational complexity. Section
\ref{sec:compare} presents the comparative study approach and results, and
Section \ref{sec:conclusion} concludes the paper with suggestions for
real-world use of these algorithms and future research directions.

\section{Test Systems and Data}\label{sec:data}
There are many types of data that can be collected from power grid systems.
This paper employs data collected by Phasor Measurement Units (PMUs).
PMUs, also called \emph{synchrophasors}, collect data synchronously across
the power grid, providing an ``online'' data stream. These units are deployed
to many systems throughout the grid and are time synchronized so that
measurements taken at different locations can be correlated together. This
work focuses on phasor data, the angle of the terminal voltage. These data
are collected at the millisecond resolution, 100 samples per second, and
afford a picture of voltage angle oscillation at each generator over time.
For each generator, a time series of phasor values is collected,
$\vec{\delta}_i = \tup{\delta_i^{(1)}, \delta_i^{(2)}, \ldots,
\delta_i^{(m)}}$, where $\delta_i^{(j)}$ is the phasor value of generator $i$
at the $j^{th}$ time step. In both test systems, data are collected for 3.8
seconds, $m=380$, but, for generality, the symbolic parameter $m$ is used
throughout the paper. In future work, a study is planned to determine how
many samples are needed to yield a useful reduced order model.


\subsection{Evaluated Test Systems}\label{sec:test}

\begin{figure}[h]
  \centering
  \includegraphics[width=0.9\linewidth]{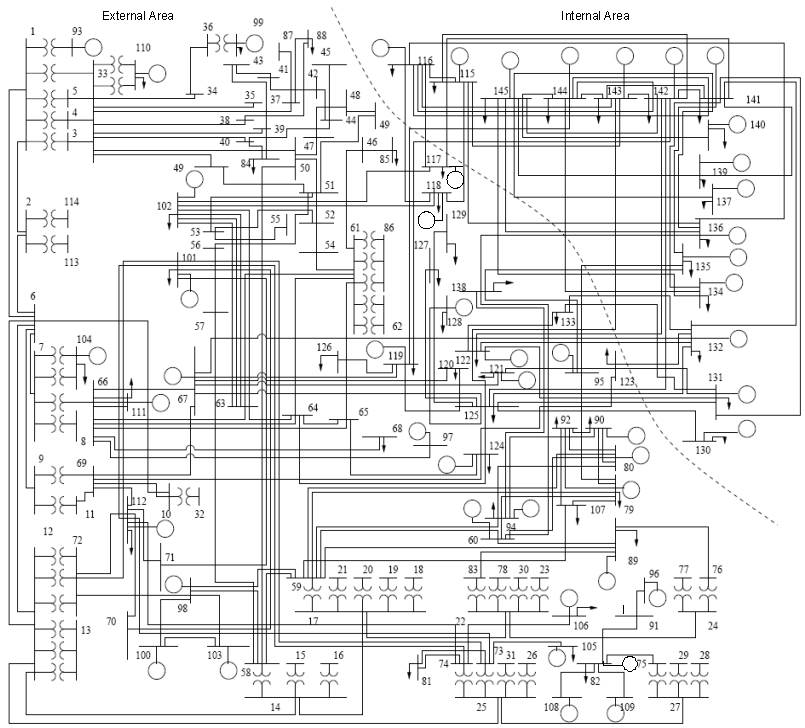}
  \caption{The IEEE 50 generator system, $S_1$. In this image, the circles represent the
  50 generators in the system, and the buses are numbered 1 through 145.}\label{fig:IEEE50}
\end{figure}

\begin{figure}[h]
  \centering
  \includegraphics[width=0.8\linewidth]{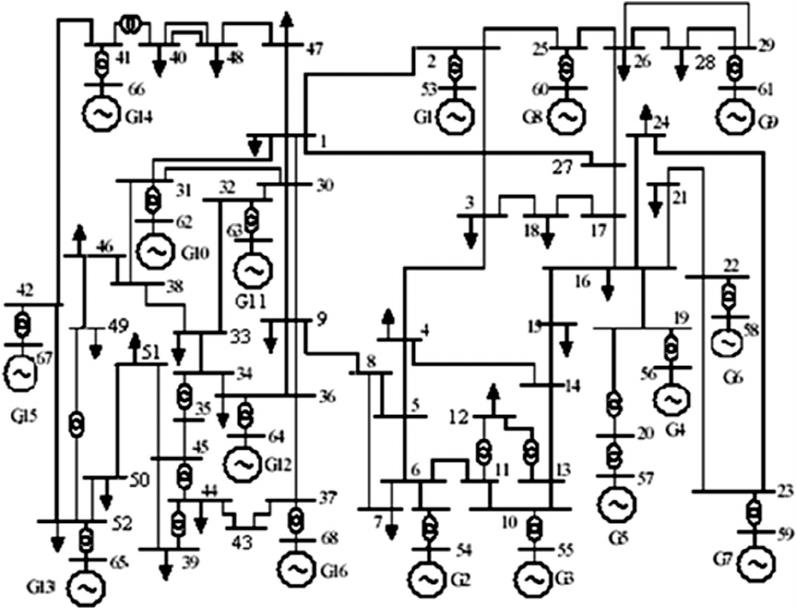}
  \caption{The IEEE 16 generator system, $S_2$. Again, the circles are generators,
  labeled G1 through G16, and the buses are numbered 1 through 68.}\label{fig:IEEE16}
\end{figure}

Two small model systems are used for method validation. Specifically, the
IEEE 145-bus, 50-generator system \cite{canizares_linear_2004}
(\fig{fig:IEEE50}) referred to as \emph{system $S_1$}, and the IEEE 68-bus,
16-generator system, \emph{system $S_2$} (\fig{fig:IEEE16}). In large power
networks, such as the Eastern or Western Interconnects in the United States,
individual power companies only control small areas, known as \emph{service
territories}. All of these small areas are interconnected to form the entire
grid. Each power company considers the generators and buses in their service
territory to be their \emph{internal system}, while the remaining generators
and buses are \emph{external}. Companies prefer to model their own internal
generators fully and use model reduction to determine a simpler approximation
for the external system as it typically is much larger. In $S_1$, there are
$I_1 = 16$ internal system generators and $E_1 = 34$ external generators,
while in $S_2$, there are $I_2 = 7$ internal and $E_2 = 9$ external. In
$S_1$, generator 37 at Bus 130 in the internal area is chosen as the
reference machine, meaning it is treated as if its phasor angle is always 0.
All other machines' phasor values are measured as deviations from the
reference machine. In $S_2$, generator 16 is used as the reference.


\begin{figure}[h]
  \centering
  \includegraphics[width=0.4\linewidth]{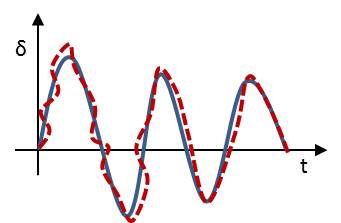}
  \caption{An illustration of the difference between the classical model response
  (solid blue line) and the high-order model response (dotted red line).}\label{fig:classicvshigh}
\end{figure}

All generators are modeled using the classical model for machine dynamics
with a second-order swing equation. In power systems, following a
disturbance, some state variables decay fast, called \emph{fast variables},
such as those from excitation systems. Meanwhile, others, known as \emph{slow
variables}, decay slowly, e.g., rotor angle and speed. Oscillations of slow
variables are determined by machine inertias and are well captured by the
classical (second-order) model \cite{ChowAccariPrice}. Moreover, slow
variables dominate power oscillations in the power system. Therefore, the
classical model is used in this work. Figure \ref{fig:classicvshigh}
illustrates the difference between the classical model response (solid blue
line) and high-order model response (dotted red line).

For the tests on $S_1$, simulated PMU data sets are created using $F_1=5$
different three-phase, short circuit faults within the system. The faults
last for 60 ms. Then, the line is tripped to clear the fault. In $S_2$, $F_2
= 3$ faults are simulated. Post-fault oscillations of the phasor values for
the external system generators are recorded. The next section describes the
methods used for model reduction to determine sets of representative
generators.

\section{Identifying Representative Generators}\label{sec:methods}

This section describes four methods for identifying representative
generators. These representative generators are meant to exhibit different
types of dynamic behavior and will be used in reduced model simulations
(described in Section \ref{sec:reduce}). Recalling the workflow given in
Figure \ref{fig:workflow} that identifies each model reduction method, this
section describes their basic details. For full descriptions of these
methods, refer to other detailed papers \cite{hogan_towards_2013,
wang_measurement-based_2012}. The computational complexity of each method is
also discussed. Table \ref{tab:clustering} summarizes the four methods in
terms of their parameters and computational complexity. Runtimes for all
methods on a standard desktop computer are less than one second for both test
systems considered.

\begin{table}
\begin{center}
\begin{tabular}{|p{0.2\textwidth}|p{0.3\textwidth}|p{0.3\textwidth}|}
\hline
\textbf{Method}                   & \textbf{Parameters} & \textbf{Computational} \newline \textbf{complexity}\\\hline \hline
SVD                               & $k$ & $O(n^2m+nm^2+m^3+nk)$ \\\hline
$k$-means                         & $k$  & $O(nk^2i)$ \\\hline
Recursive spectral bipartitioning & $k$; $\ell$;  ``median'' or ``zero''; ``size'', ``sum'', or ``avg''& $O((k-1)n^3)$ \\\hline
Spectral \newline clustering & $k$; $\ell$ & $O(n^3)$ \\\hline
\end{tabular}
\end{center}
\caption{Summary of model reduction methods investigated. Refer to each
corresponding section for a detailed description of the parameters.}\label{tab:clustering}
\end{table}

\subsection{Singular Value Decomposition (SVD)}\label{sec:SVD}
SVD's goal in the power grid model reduction setting is to find a subset of
external generators, known as \emph{representative generators}, whose dynamic
responses, the $\vec{\delta}_i$ vectors, are as close to orthogonal as
possible. The more orthogonal the representative generators are, the larger
their span. Therefore, there is a better chance that the linear span of these
representative vectors will contain the dynamic response vectors for the
remainder of the generators.

SVD goes far beyond this application and is a general method of matrix
factorization into two unitary matrices and one diagonal matrix
\cite{GeJ1998}. Given an initial $m \times n$ matrix, $A$, the SVD method
factorizes $A$ into a product of three matrices, $A = U \Sigma V^*$, where
$U$ ($m \times m$) and $V$ ($n \times n$) are unitary and $\Sigma$ is an $m
\times n$ diagonal matrix. The columns of $U\Sigma$ are the \emph{principal
components} of $A$, while the diagonal values of $\Sigma$ are \emph{singular
values}. Singular values are often used to determine how many principal
components to choose because smaller singular values tend to contribute only
noise to the principal components.
SVD is used across many different domains, including audio verification
\cite{Sahidullah20161} and evolutionary genomics \cite{genomicsSVD}. This
work focuses on the algorithm described in
\cite{wang_measurement-based_2012}, where SVD is used for power grid model
reduction.

Recall that for each external generator, $i$, a time series of phasor values
is collected from a PMU and normalized to form the vector $\vec{\delta}_i =
\tup{\delta_i^{(1)}, \delta_i^{(2)}, \ldots, \delta_i^{(m)}}$, where
$\delta_i^{(j)}$ is the normalized phasor value of generator $i$ at the
$j^{th}$ time step, and $m$ is the number of time steps. This vector
represents the dynamics of generator $i$ following a disturbance. The
normalization is done in a standard way by subtracting the mean and dividing
by the standard deviation for each $\vec{\delta}_i$ separately. Let $n=E_i$
be the number of external generators (in system $S_i$) and define the matrix
$\delta$ to have $\vec{\delta}_i$ as column vectors.
\[ \delta = \left[ \begin{array}{cccc}
                   \vertbar       & \vertbar       &        & \vertbar\\
                   \vec{\delta}_1 & \vec{\delta}_2 & \cdots &\vec{\delta}_n\\
                   \vertbar       & \vertbar       &        & \vertbar
\end{array}\right] \]

Given this formulation, an SVD is performed on the matrix $\delta$, writing
$\delta = U \Sigma V^*$. The first $k$ principal components are found by
taking the $k \times k$ submatrix of $\Sigma$ that has the $k$ largest
singular values, denoted $\Sigma_k$, along with the corresponding $k$ columns
of $U$, denoted $U_k$. It can be written as
\[ U_k \Sigma_k = X_k = \left[ \begin{array}{cccc}
                   \vertbar       & \vertbar       &        & \vertbar\\
                   \vec{x}_1      & \vec{x}_2      & \cdots &\vec{x}_k\\
                   \vertbar       & \vertbar       &        & \vertbar
                   \end{array}\right] \]
where $\vec{x}_i$ is the $i^{th}$ principal component.
%
The computational complexity of computing the principal components is $O(n^2
m + n m^2 + m^3)$ \cite{GoGVaC1996}. For both test systems considered in this
paper, the number of time steps, $m$, is much larger than the number of
generators, $n$, so the $m^3$ term will dominate. However, this generally may
not be the case, particularly in an online process where few time steps from
the recent history on a large system are used.

\begin{figure}
    \centering
    \begin{subfigure}[b]{0.48\textwidth}
        \includegraphics[width=0.95\textwidth]{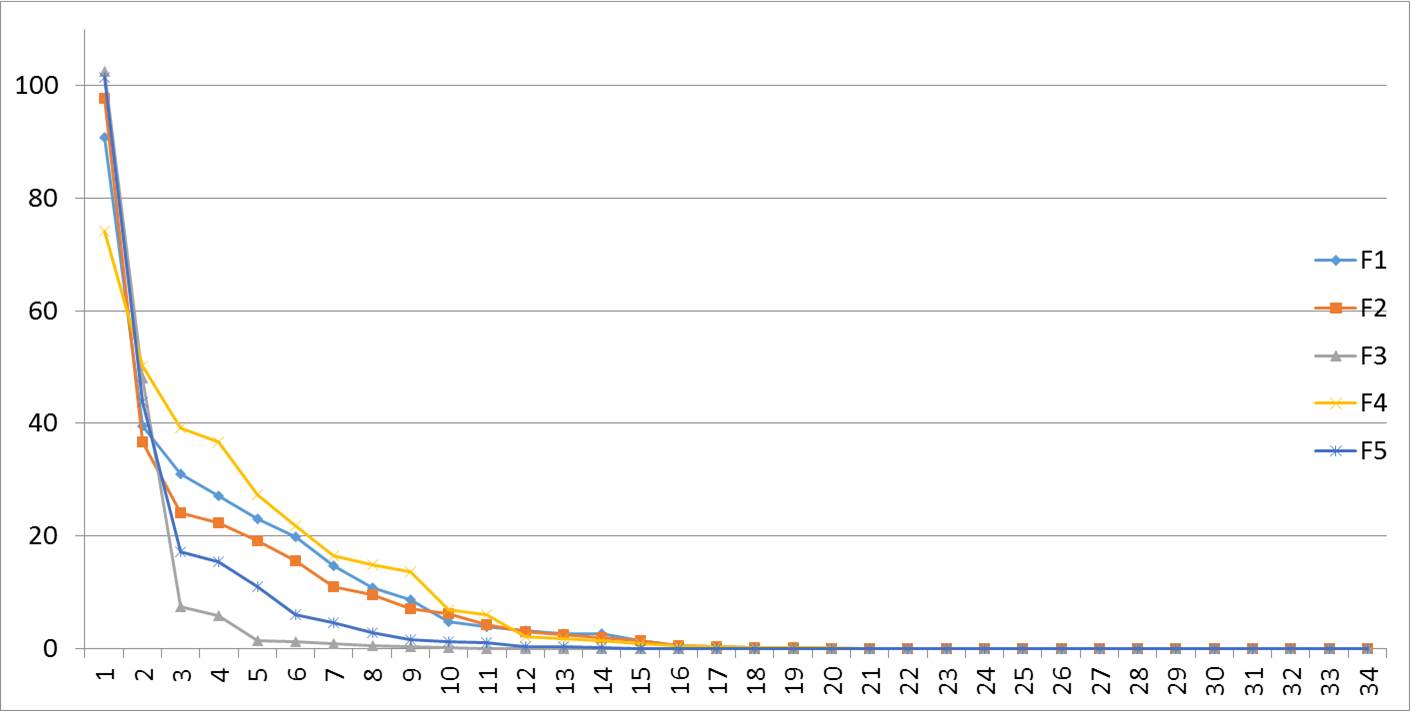}
        \caption{Singular values for all five fault scenarios in the IEEE 50 system, system $S_1$.}
        \label{fig:IEEE50_svd}
    \end{subfigure}
    \hspace{0.05\textwidth}
    \begin{subfigure}[b]{0.42\textwidth}
        \includegraphics[width=0.90\textwidth]{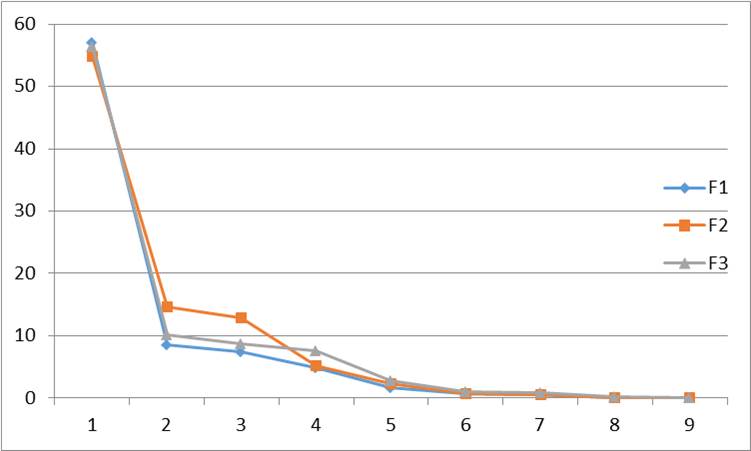}
        \caption{Singular values for all three fault scenarios in the IEEE 16 system, system $S_2$.}
        \label{fig:IEEE16_svd}
    \end{subfigure}
    \caption{}\label{fig:singulars}
\end{figure}

The two plots in Figure \ref{fig:singulars} show the singular values
calculated for systems $S_1$ (a) and $S_2$ (b). Because smaller singular
values can contribute noise to the principal components, these plots are used
to pick the cutoffs of 4 to 15 in system $S_1$ and 3 to 6 in system $S_2$.

Once the principal components, $X_k$, are computed, the similarity between
$\vec{\delta}_i$s  and $\vec{x}_j$s is analyzed. Namely, for each $\vec{x}_j$
find the $\vec{\delta}_i$ closest in the Euclidean distance and choose that
$\vec{\delta}_i$ to be one of the representative generators. This yields the
$k$ vectors, $\vec{\delta}_i$, with the highest similarity to an $\vec{x}_j$,
forming the set of $k$ representative generators. The complexity of this step
is $O(nk)$ because each of the $n$ phasor vectors must be compared to each
$k$ principal components.

\subsection{$k$-Means}
Because the PMU data are vector data and can be thought of as points in
$\R^m$ for some $m$, $k$-means clustering is a natural choice for comparison.
$k$-means is a standard recursive clustering technique for data in $\R^m$
\cite{selim_k-means-type_1984} and is widely used
\cite{honarkhah_stochastic_2010, LiDWuX09}. In each recursion step, the
algorithm computes centroids of each current cluster and then reassigns
points to the cluster whose centroid is closest to it. An initialization step
is needed to choose the initial centroids, which is often done by choosing
$k$ points randomly from the original data set. The algorithm then runs as
follows: assign each data point to the cluster whose randomly chosen centroid
is closest, recompute centroids of the current clusters, reassign points to
clusters based on distance to new centroids, recompute centroids, etc. This
is repeated for some predetermined number of iterations, or until the
clusters do not change and the algorithm has converged.

One problem with $k$-means is that there is no guarantee it will terminate in
a globally optimal clustering. $k$-means uses gradient descent, or a similar
optimization algorithm, at each step. These are well known to have the
possibility of getting stuck in a locally optimal clustering. Because the
initialization step is done randomly, there can be multiple clusterings from
the same input data, all being local optima. This problem can be mitigated by
running $k$-means clustering multiple times with different random
initializations and choosing the resulting clustering that minimizes an
objective function. The chosen objective function is the \emph{residual sum
of squares} defined as
\[ RSS = \sum_{i=1}^n ||\vec{\delta_i} - \mu(\vec{\delta}_i)||^2 \]
where $\mu(\vec{\delta}_i)$ denotes the centroid for the cluster that
contains $\vec{\delta}_i$ and the norm is taken to be the Euclidean norm. We
use the scikit-learn python implementation of $k$-means, which runs a default
of 10 times and chooses the clustering with the smallest $RSS$ value.

This local optimum problem is also encountered in the spectral clustering
method, which uses $k$-means. However, it is not an issue for SVD or
recursive spectral bipartitioning. The advantage of $k$-means, however, is
that it is the fastest of the algorithms considered in this paper. The
complexity of $k$-means is $O(nk^2 i)$, where $i$ is the number of iterations
required to converge. As $k$ will be less than $n$, this is faster than the
$O(n^3)$ graph clustering algorithms.

\subsection{Graph Clustering}
Recursive spectral bipartitioning, and spectral clustering methods are not
new, but they have not previously been used to perform power grid model
reduction. We have made modifications that will allow for fine tuning by an
operator.

First, the graph construction method from PMU data must be described. When
setting up a graph clustering problem, the graph vertices are chosen as the
objects being clustered (in this case, the generators), while the edges will
indicate an amount of similarity between vertices.
For each generator, consider its phasor value data vector, $\vec{\delta}_i
\in \R^m$, where $m$ is the number of time steps recorded, and calculate a
distance matrix $D = (d_{ij})_{i,j=1}^n$. The entries in $D$ are given by the
Euclidean distance between $\vec{\delta}_i$ and $\vec{\delta}_j$, $d_{ij} =
||\vec{\delta}_i - \vec{\delta}_j||_2$. Once this matrix has been created, an
$\ell$-\emph{nearest-neighbor graph} is formed by connecting each generator
(vertex) to
its $\ell$ closest generators. Alternate distances and graph constructions
also can be used (see \cite{hogan_towards_2013} for more details).

Both types of graph clustering will use spectral (eigenvalue) properties of
the weighted Laplacian matrix associated with the graph. The \emph{weighted
Laplacian}, $L = (L_{ij})_{i,j=1}^n$, is defined as follows:
\[ L_{ij} = \left\{ \begin{array}{ll}
                  \sum_{k \neq i} w_{ik} & i=j\\
                  -w_{ij} & i \neq j.
                  \end{array} \right. \]
The entries on the diagonal, $L_{ii}$, are given by the sum of all edge
weights on edges incident to vertex $i$. Off-diagonal entries, $L_{ij}$, are
equal to the negative weight on edge $e_{ij}$. If an edge is absent, it is
treated as an edge of weight zero. An edge's weight is defined as the
similarity score between the endpoint vertices based on their distance (high
distance means low similarity, and low distance represents high similarity).
In particular, let $w_{ij} = e^{-(d_{ij}^2/2)}$ be the Gaussian similarity
between generators $i$ and $j$. Other similarity functions, also known as
\emph{kernels}, may be used, but the Gaussian function is fairly standard
\cite{luxburg_tutorial_2007}.

Creating the graph and the Laplacian matrix has complexity $O(n^2)$. Begin by
calculating all pairs of distances between the $n$ generators. This will be
$\frac{n(n-1)}{2}$ distance calculations and dominates the complexity. Then,
to construct the $\ell$-nearest neighbor graph, for each vertex, consider all
other vertices and connect to the $\ell$ closest. This requires sorting the
sets of neighbors for each of the $n$ vertices. Therefore, creating an
$\ell$-nearest neighbor graph can be done in $O(n \log n)$ time after the
distances have been computed.

\subsubsection{Recursive Spectral Bipartitioning}\label{sec:fiedler}
The most basic type of spectral graph clustering or partitioning is recursive
spectral bipartitioning \cite{Pothen:1990:PSM:84514.84521}. This algorithm
uses the eigenvector for the second smallest eigenvalue of the weighted
Laplacian matrix. Clearly, as each row of the Laplacian matrix sums to zero,
there is a zero eigenvalue. It is not difficult to show that $L$ is positive
semidefinite. Thus, zero is, in fact, the smallest eigenvalue, and its
multiplicity is the number of connected components in the graph. The second
smallest eigenvalue is the \emph{algebraic connectivity} of the graph, and
its associated eigenvector, commonly called the \emph{Fiedler vector} after
Miroslav Fiedler who first defined the theory of algebraic connectivity and
its relation to graph partitioning \cite{fiedler1, fiedler2}, has properties
that define a partition of the graph vertices into two groups. The Fiedler
vector contains positive, negative, and zero values. By partitioning the
associated vertices into two sets --- one where the value in the Fiedler
vector is negative and the other in which it is positive (splitting the zero
values among both or putting them all in one of the two sets) --- a graph
partition is obtained that minimizes the sum of the edge weights between the
two partitions, and where both subgraphs are connected \cite{fiedler1,
fiedler2}. Traditionally, the graph vertices are partitioned into those with
positive values and others with negative values because of the property that
both subgraphs are connected. However, this can lead to unbalanced partitions
because there is no guarantee that half of the vertices will have positive
values while the other half will be negative. To construct more balanced
partitions, the split can be made based on the median of the Fiedler vector.
This choice only guarantees that one of the two induced subgraphs is
connected. However, since the partition will be recursively continued, having
disconnected induced subgraphs should not create problems because they will
have the opportunity to split eventually. Both of these splitting methods
have been tested and are reported as part of the comparison in Section
\ref{sec:compare}. In particular, splitting at zero often gives degenerate
partitions into fewer than $k$ sets.

Using the Fiedler vector to partition the graph vertices into two disjoint
sets is only the first step. A second degree of freedom in this method is how
to continue the recursive partitioning. In traditional recursive spectral
bipartitioning, both of the sets are further partitioned using the Laplacian
of the subgraph induced by each set of vertices. If this process of
repeatedly splitting each set is continued for $N$ steps, it will yield a
partition with $2^N$ sets. A more targeted approach to the recursive
splitting is explored to achieve any number of sets rather than just powers
of 2. Instead of arbitrarily splitting each set of the partition into two at
each step, a search is performed among all current sets for one that is the
least \emph{tight}. We define tight as defined as either the \emph{sum} of
all pairwise distances in that set, the \emph{average} of all pairwise
distances, or simply the \emph{size} of the set. These three possible
tightness schemes are considered, in addition to the two methods for
splitting the Fiedler vector (zero or median). All of the results are
summarized in \sect{sec:compare}. In the remainder of this paper, recursive
spectral bipartitioning may be referred to as \emph{Fiedler partitioning}
because of the prominent use of the Fiedler vector.

The computational complexity of recursive spectral bipartitioning is
dominated by the Fiedler vector calculation which has complexity $O(n^3)$.
This must be done many times, each time a set is split into two. Therefore,
complexity is $O((k-1)n^3)$.

\subsubsection{Spectral Clustering}
\begin{figure*}
  \centering
  \includegraphics[width=0.9\textwidth]{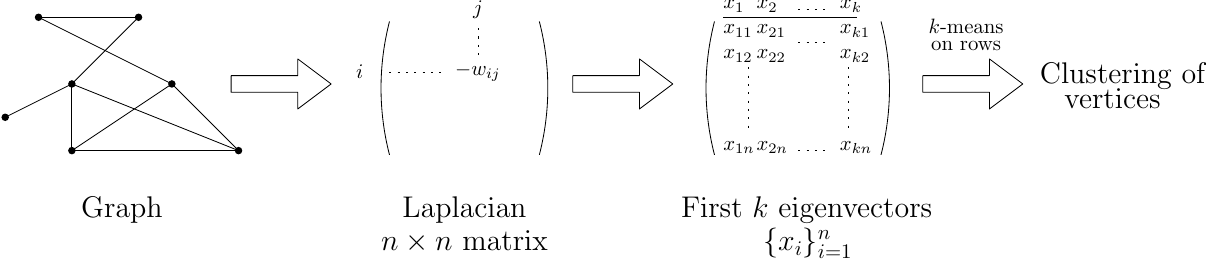}
  \caption{An illustration of the steps in spectral clustering.}\label{fig:spec-clust}
\end{figure*}

For general spectral clustering, more than just one eigenvector of the
weighted Laplacian is used. Instead of using only the Fiedler vector, the
first $k$ eigenvectors are considered. Each entry of an eigenvector
corresponds to a vertex in the graph from which the weighted Laplacian was
formed. An $n \times k$ matrix can be formed where the columns are the first
$k$ eigenvectors. Each row can be thought of as a new vector representation
for each of the vertices. $k$-means is then used to cluster these new vector
representations, thereby clustering the vertices themselves. The pipeline
illustration in \fig{fig:spec-clust} shows this sequence. In this case,
eigenvectors are only calculated once, at $O(n^3)$ complexity, followed by a
single $k$-means calculation. As the $k$-means complexity is less than $n^3$,
the total complexity of spectral clustering is dominated by the $n^3$ term.
For a more in-depth discussion of spectral clustering, including other
variants not investigated here, see \cite{luxburg_tutorial_2007}.

\subsection{Choosing Cluster Representatives}
Both $k$-means and the graph methods produce a partition of the generators
into clusters of similar dynamic behavior. However, in the context of model
reduction, we need a set of \emph{representative generators}, as produced by
the SVD method. This is achieved by choosing one representative generator
from each cluster: the medoid.

For each cluster, $C_i = \{ \delta_{i_j} \}_{j = 1}^{|C_i|}$, it is
preferable to choose the average time series, the true centroid:
\[ \delta_{C_i} = \tup{ \frac{\sum_{j = 1}^{|C_i|} \delta_{i_j}^{(k)}}{|C_i|} }_{k=1}^m. \]
However, this true centroid is unlikely to correspond exactly to the time
series of phasor values for a generator in the given cluster. Instead, the
generator whose time series is \emph{closest} to this centroid in the
Euclidean distance is chosen. This element closest to the true centroid is
known as the \emph{medoid} of the cluster.


\section{From Representative Generators to a \\Reduced Model Simulation}\label{sec:reduce}

Once a set, $C$, of representative generators has been produced from any one
of the methods described in Section \ref{sec:methods}, a reduced model
simulation can be performed. For each non-representative generator,
$\vec{\delta}_i \not\in C$, coefficients $\alpha_j \in \R$ are found via
regression such that $||\vec{\delta}_i - \sum_{\vec{\delta}_j \in C}
\alpha_j\vec{\delta}_j||_2$ is minimized. In other words, the phasor values
for each non-representative generator are approximated as a linear
combination of the phasor values for the representative generators. Then,
only the set of representative generators are simulated, by solving power
flow equations, and responses for the remaining generators are approximated
using the same linear combinations of the responses for the representative
generators.

\section{Comparative Study Approach and Results}\label{sec:compare}

In this section, a measure that quantifies the amount of error between a
reduced model and the full system is defined. In addition, an overview of
performance profiles, the method for comparing the error in reduction methods
across multiple scenarios, is provided. Finally, performance profiles are
used to compare the reduction methods described in the previous section.

\subsection{Measures for Comparison}\label{sec:measures}

To judge the error of a particular model reduction method, $M$, for test
system $S_x$ ($x=1, 2$) under fault scenario $1 \leq u \leq F_x$, first the
full system of external and internal generators ($E_x$ external + $I_x$
internal) is simulated and then the reduced system ($r$ representative
external + $I_x$ internal, depending on the reduction ratio $r$). Recall that
the internal system represents the set of generators and buses owned by a
particular power company of interest and are not reduced, while the external
machines are interconnected but owned by others. Responses, or \emph{phasor
values}, of the $I_x$ internal generators are recorded during both full and
reduced simulations. Let $\delta_{u,i}^f(t)$ be the phasor value response of
the $i^{th}$ internal generator in the full simulation at time $t$ following
fault scenario $u$ and $\delta_{u,i}^{M,r}(t)$ be the same for the reduced
model. Notice that $\delta_{u,i}^f$ does not have the $M$ superscript because
it is independent of any reduced model. Then, for each internal system
generator, define the following metric to measure the mismatch of response
curves of the full and reduced systems:
\begin{align}
  J_{u, r}^{M}(i) &= \frac{1}{t_2-t_1} \int_{t_1}^{t_2} \left| \delta_{u,i}^{M,r}(t) - \delta_{u,i}^f(t) \right| dt \label{eq:Jai}
\end{align}
which is the $L_1$ norm between vectors $\delta_{u,i}^{M,r}$ and
$\delta_{u,i}^f$.
For example, $J_{3, 7}^{\text{SVD}}(38)$ for system $S_1$ is the error on
internal generator $i=38$ for fault scenario $u=3$ using the $M=$ SVD model
reduction method to $r=7$ generators. While using the $L_2$ norm or some
other $L_p$ norm is an option, the results are nearly identical in our test
cases. Hence, only one is presented.

To simplify the comparisons define
\[ J_{u, r}^{M} = \frac{1}{I_x} \sum_{i \in I_x} J_{u, r}^{M}(i) \]
to be the average $J_{u, r}^{M}(i)$ 
values over all internal generators. Of note, there is a slight abuse of
notation in the sum over $i \in I_x$. Here, we use $I_x$ to mean both the set
of internal generators and the number of internal generators, as it was
originally defined.

%

\subsection{Performance Profiles}\label{sec:perfprof}

The comparison method use, \emph{perfprof} (for ``performance profile'')
\cite{DiNHiN2013, DoEMoJ2002, HiDHiN2005}, comparatively plots the
performance of different algorithms against each other. This type of analysis
is typically used when comparing the runtimes of multiple algorithms against
each other, preferring low runtime over high. This strategy is adopted here
because the premise is the same: to choose the reduction method that most
often across multiple comparable tests (fault scenarios) has smallest error
value. Comparisons will only be made within the same reduction ratio, $r$,
and within the same test system. Once a system ($x=$ 1 or 2) and an $r$ have
been chosen, there are nine methods to compare across $F_x$ fault scenarios,
which are called \emph{tests}. Table \ref{tab:ex_compare} summarizes these
values for the example where $x=2$ and $r=5$.

\begin{table}[h]
\begin{center}
\begin{tabular}{l|lll}
  $M$ & Test 1 & Test 2 & Test 3 \\\hline
  SVD & $J_{1,5}^{\text{SVD}}$ & $J_{2,5}^{\text{SVD}}$ & $J_{3,5}^{\text{SVD}}$\\
  $k$-means & $J_{1,5}^{k\text{-means}}$ & $J_{2,5}^{k\text{-means}}$ & $J_{3,5}^{k\text{-means}}$\\
  Fiedler, zero, sum & $J_{1,5}^{\text{Fzsu}}$ & $J_{2,5}^{\text{Fzsu}}$ & $J_{3,5}^{\text{Fzsu}}$\\
  Fiedler, zero, avg & $J_{1,5}^{\text{Fza}}$ & $J_{2,5}^{\text{Fza}}$ & $J_{3,5}^{\text{Fza}}$\\
  Fiedler, zero, size & $J_{1,5}^{\text{Fzsi}}$ & $J_{2,5}^{\text{Fzsi}}$ & $J_{3,5}^{\text{Fzsi}}$\\
  Fiedler, mid, sum & $J_{1,5}^{\text{Fmsu}}$ & $J_{2,5}^{\text{Fmsu}}$ & $J_{3,5}^{\text{Fmsu}}$\\
  Fiedler, mid, avg & $J_{1,5}^{\text{Fma}}$ & $J_{2,5}^{\text{Fma}}$ & $J_{3,5}^{\text{Fma}}$\\
  Fiedler, mid, size & $J_{1,5}^{\text{Fmsi}}$ & $J_{2,5}^{\text{Fmsi}}$ & $J_{3,5}^{\text{Fmsi}}$\\
  spectral & $J_{1,5}^{\text{spec}}$ & $J_{2,5}^{\text{spec}}$ & $J_{3,5}^{\text{spec}}$\\
\end{tabular}
\end{center}
\caption{Summary of the values compared for system $S_2$ and reduction ratio $r=5$.}\label{tab:ex_compare}
\end{table}

Note that some of the methods did not return a reduced model for all $r$
values in a given fault scenario, so some of these data may be missing. This
is because the clustering may have degenerated into more than $r$ clusters.
For example, if the graph clustering resulted in $r$ clusters where one of
the clusters consisted of only $v$ isolated vertices, the algorithm treats it
as $r+(v-1)$ clusters. If this is the case, we let the corresponding
$J_{u,r}^M$ measure be a value larger than any of the other values returned
in the table. This indicates that it did a ``bad job'' at that particular
reduction ratio in that test or scenario. Too many of these degenerate
reductions reflects poorly on the method and will lower the perfprof score,
described next, as its error will always be much higher than the lowest
error.

\begin{figure}[h]
  \begin{center}
  \includegraphics[width=0.5\textwidth]{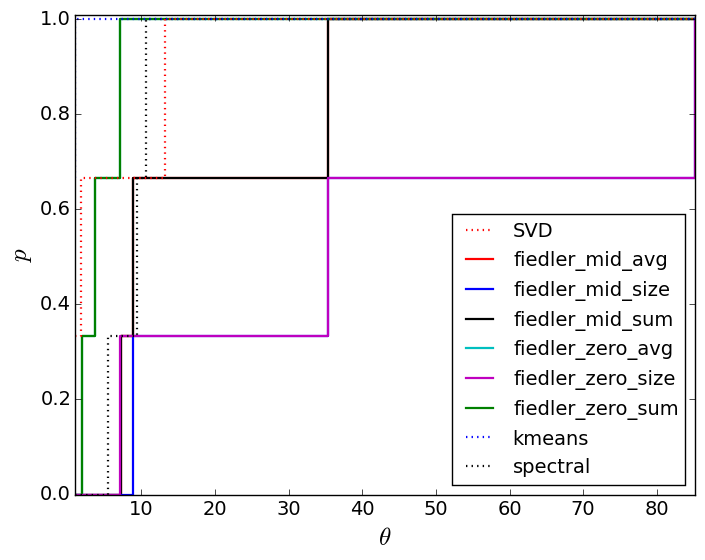}
  \end{center}
  \caption{The performance profile for system $S_2$ and reduction ratio $r=5$.
  A point $(\theta, p)$ for method $M$ means that $J_{u,5}^{M}$ is within a factor of
$\theta$ of the best $J_{u}$ in proportion $p$ of the three fault scenarios.}\label{fig:S2r5Ja}
\end{figure}

The performance profile method produces a plot with a \emph{tolerance
factor}, $\theta$, on the $x$ axis and a \emph{proportion} of the tests,
$0\leq p\leq 1$, on the $y$ axis. Each of the model reduction methods
corresponds to a staircase-shaped curve in the plot. Continuing to use Table
\ref{tab:ex_compare} as an example, a point $(\theta, p)$ for method $M$ in
this example means that $J_{u,5}^{M}$ is within a factor of $\theta$ of the
best $J_{u,5}$ in proportion $p$ of the three tests. It is likely that the
best method is different for each $u$, but the $J_{u,5}^M$ measurement
produced by this model reduction method is within some factor of whatever the
best is for each test. In particular, a point $(1, p)$ means that
$J_{u,5}^{M}$ is the optimal value (i.e., within a factor of 1 of the best)
in proportion $p$ of the tests, and a point $(\theta, 1)$ denotes that
$J_{u,5}^{M}$ is always (proportion $p=1$) within a factor of $\theta$ of the
optimal value. Figure \ref{fig:S2r5Ja} contains the perfprof plot for this
specific case. The method Fiedler-zero-size (purple line) passes through the
point $\sim$$(35, 0.66)$, meaning that for roughly 66\% of the tests ($u$
values), the Fiedler-zero-size $J_{u,5}^{Fzsi}$ value is within a factor of
35 of the best $J_{u,5}^{M}$. This factor of 35 may sound high, but these
measures range between $\sim$0.002 and 1, making the maximum possible factor
around 500.

Given the perfprof plot for an $S_x$ and $r$, observe that a line which stays
close to the $p$ axis the longest and reaches $p=1$ first is optimal.
However, there may not be a single method that achieves both of these
objectives. Instead, we posit that generally higher and further left is
better. Therefore, much like judging the accuracy of a binary classification
algorithm by area under an ROC curve, we compare the accuracy of model
reduction methods using area under its perfprof curve. Unlike ROC curves, the
$x$ axis ($\theta$) is not bound between 0 and 1. Instead, the maximum
$\theta$ value is the maximum ratio between any two accuracy measures for the
given reduction ratio. That is to say, for a given $r$, the maximum $\theta$
will be
\[ \max \theta = \max_{M_1, M_2, u} \frac{J_{u,r}^{M_1}}{J_{u,r}^{M_2}} \]
where the maximum is taken over all pairs of methods $M_1$ and $M_2$ and all
fault scenarios $u$. To normalize the area, divide by the total possible
area, which is $\max \theta$ as the $y$ axis maximum is 1. In the next
section, this perfprof comparison method is used to determine optimal model
reduction methods for each system and reduction ratio.

\subsection{Results of Comparison}\label{sec:results}

For system $S_1$, reduction ratios $4\leq r \leq 15$ are considered, and for
system $S_2$ we let $3 \leq r \leq 6$. This totals $12 + 4 = 16$ different
perfprof comparisons. Therefore, rather than including all perfprof plots,
two tables of normalized areas under the perfprof curves will be provided,
one for each test system, showing all $r$ values. In each table, values are
rounded to three significant digits. In particular all 1.000 values are
rounded to that value and are not equal to it. To reinforce the tables, some
perfprof plots will also be included.

\subsubsection{System $S_1$}
In system $S_1$, the methods Fiedler-zero-sum and Fiedler-zero-size only
return reductions for $r=4$ and Fiedler-zero-avg only for $r=4, 5, 6$.
Therefore, those three methods have the worst performance in our comparisons.
In this system, splitting the Fiedler vector at zero sometimes yields a
degenerate partition, where one part is empty. Therefore, the splitting
process reaches a stable point (nothing else can be split using the Fiedler
vector) prior to obtaining the desired number of clusters.

\begin{figure}[h]
\begin{center}
  \includegraphics[width=0.5\textwidth]{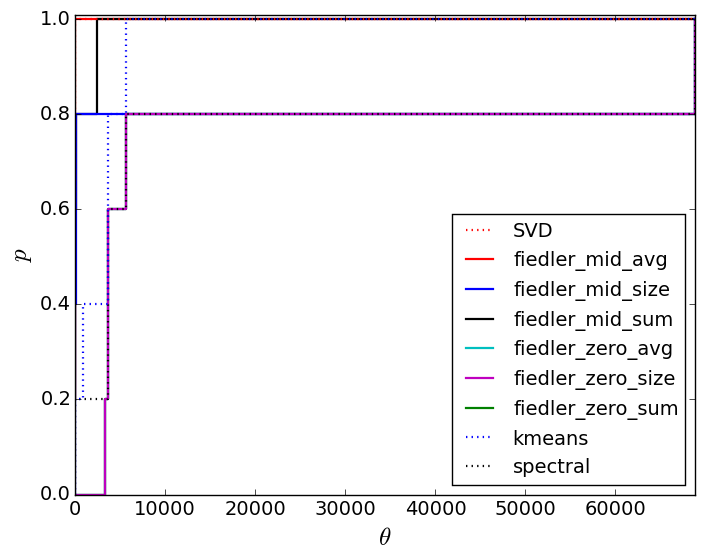}
\end{center}
\caption{Perfprof plot for system $S_1$ and $r=10$. The
curves for Fiedler-zero-avg and Fiedler-zero-sum are hidden under the purple
Fiedler-zero-size curve as all three did not return any reductions for $r=10$. A point $(\theta, p)$ for
method $M$ means that $J_{u,10}^{M}$ is within a factor of
$\theta$ of the best $J_{u,10}$ in proportion $p$ of the five fault scenarios.}
\label{fig:S1r10Ja}
\end{figure}

Table \ref{tab:S1Ja} shows the normalized area measurements for system $S_1$.
The methods are in decreasing order by their average area under the curve
over all $r$ values. It is clear that SVD dominates in this case. However,
notice that for $r=5, 6$, SVD does not yield the maximal area under the
perfprof curve. In this case, it is beat by Fiedler-mid-sum and
Fiedler-mid-size, the methods with the second and third largest average
areas. Figure \ref{fig:S1r10Ja} shows an example perfprof plot for $r=10$.
The curves for Fiedler-mid-size (solid blue) and $k$-means (dotted blue)
cross around (3000, 0.8), but the area under $k$-means is higher than that
for Fiedler-mid-size, indicating that $k$-means would be preferred between
the two for $r=10$.

\begin{table}[h]
\begin{center}
{\tiny
\begin{tabular}{l|llllllllllll|l}
Method \textbackslash~ $r$ & 4 & 5 & 6 & 7 & 8 & 9 & 10 & 11 & 12 & 13 & 14 & 15 & Avg\\  \hline
SVD	&0.989	&0.999	&0.978	&1.000	&1.000	&1.000	&1.000	&1.000	&1.000	&1.000	&1.000	&0.999	&0.997 \\
Fiedler-mid-sum	&0.699	&1.000	&1.000	&1.000	&1.000	&0.977	&0.993	&0.998	&0.800	&1.000	&0.990	&0.996	&0.954 \\
Fiedler-mid-size	&0.699	&1.000	&0.982	&0.994	&0.999	&1.000	&0.800	&1.000	&0.994	&0.994	&0.990	&0.996	&0.954 \\
Fiedler-mid-avg	&0.699	&0.975	&0.975	&1.000	&0.775	&0.991	&1.000	&1.000	&0.988	&1.000	&0.991	&0.756	&0.929 \\
Spectral	&0.979	&0.745	&0.974	&0.985	&1.000	&0.787	&0.763	&1.000	&0.990	&0.792	&0.772	&0.710	&0.875 \\
$k$-means	&0.779	&0.745	&0.965	&0.985	&1.000	&0.787	&0.960	&0.992	&0.996	&0.792	&0.772	&0.710	&0.874 \\
Fiedler-zero-avg	&0.755	&0.880	&0.911	&0.733	&0.631	&0.765	&0.753	&0.598	&0.777	&0.787	&0.763	&0.710	&0.755 \\
Fiedler-zero-sum	&0.755	&0.680	&0.711	&0.733	&0.631	&0.765	&0.753	&0.598	&0.777	&0.787	&0.763	&0.710	&0.722 \\
Fiedler-zero-size	&0.755	&0.680	&0.711	&0.733	&0.631	&0.765	&0.753	&0.598	&0.777	&0.787	&0.763	&0.710	&0.722
\end{tabular}
}
\end{center}
\caption{Relative areas under the perfprof curves for system $S_1$, rows
ordered by average value.}\label{tab:S1Ja}
\end{table}

\begin{figure}[h]
\centering
    \begin{subfigure}[b]{0.45\textwidth}
      \includegraphics[width=\linewidth]{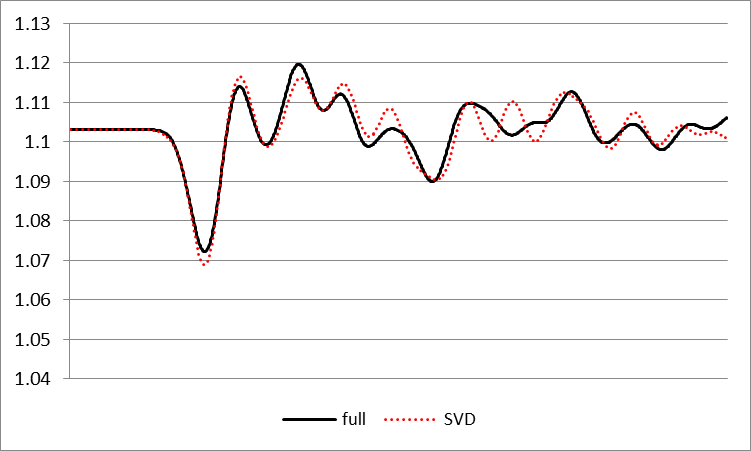}
      \caption{Full model vs. SVD}
    \end{subfigure}
    \hspace{0.05\textwidth}
    \begin{subfigure}[b]{0.45\textwidth}
      \includegraphics[width=\linewidth]{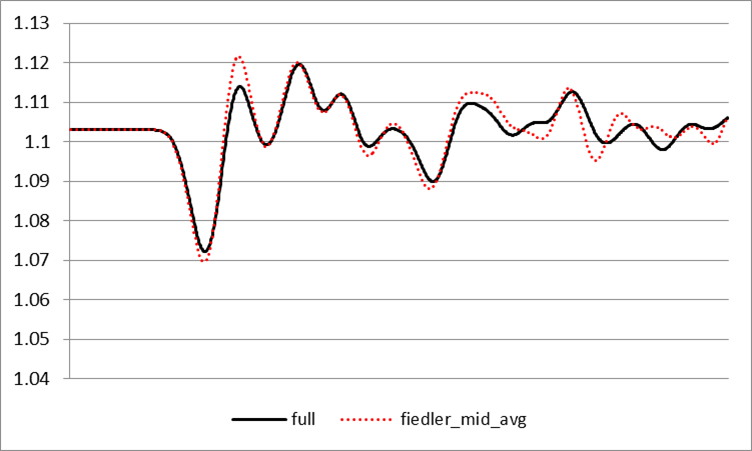}
      \caption{Full model vs. Fiedler-mid-avg}
    \end{subfigure}
    \\
    \begin{subfigure}[b]{0.45\textwidth}
      \includegraphics[width=\linewidth]{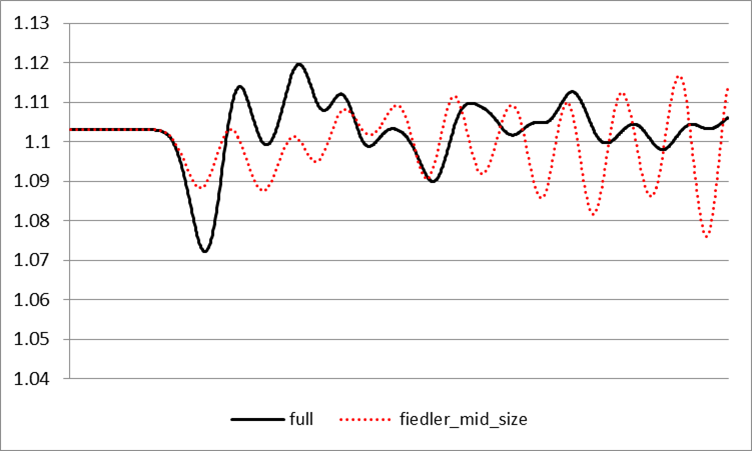}
      \caption{Full model vs. Fiedler-mid-size}
    \end{subfigure}
    \hspace{0.05\textwidth}
    \begin{subfigure}[b]{0.45\textwidth}
      \includegraphics[width=\linewidth]{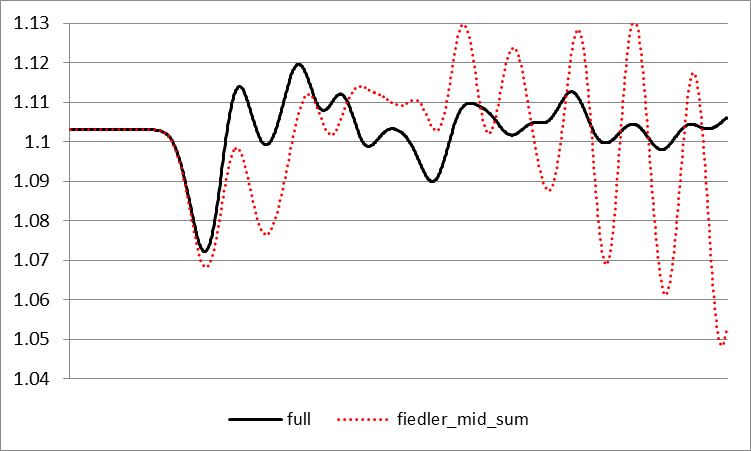}
      \caption{Full model vs. Fiedler-mid-sum}
    \end{subfigure}
    \caption{PMU traces for generator 45 in system $S_1$ with the first fault scenario under the full model
    (no reduction) and three separate reduced models for $r=10$.}\label{fig:pmu_45}
\end{figure}

These perfprof area comparisons may seem far removed from the actual PMU
phasor values. To further illustrate the comparisons, some PMU traces for
full and reduced systems under four different methods are produced. Figure
\ref{fig:pmu_45} contains four PMU response comparisons for internal system
generator \#45 following fault scenario 1. In all four subfigures, the solid
black line represents the phasor values over time for the full system without
any model reduction, and the dotted red line represents the phasor values
simulated using the indicated model reduction for $r=10$. The $r=10$ column
in Table \ref{tab:S1Ja} indicates that SVD and Fiedler-mid-avg have the best
reductions fairly consistently across all five fault scenarios.
Fiedler-mid-sum scores very high, and Fiedler-mid-size does not perform as
well. One might draw the same conclusions from looking at the PMU traces,
though it may not be as clear-cut as it is when looking at the perfprof
areas. In particular, Fiedler-mid-sum appears to perform very poorly, but its
perfprof area remains quite high. Of course, this is only one of 16 internal
generators and only a single fault scenario, so the Fiedler-mid-sum method
must have performed much better for other generators.

\begin{figure}[h]
\centering
    \begin{subfigure}[b]{0.45\textwidth}
      \includegraphics[width=\linewidth]{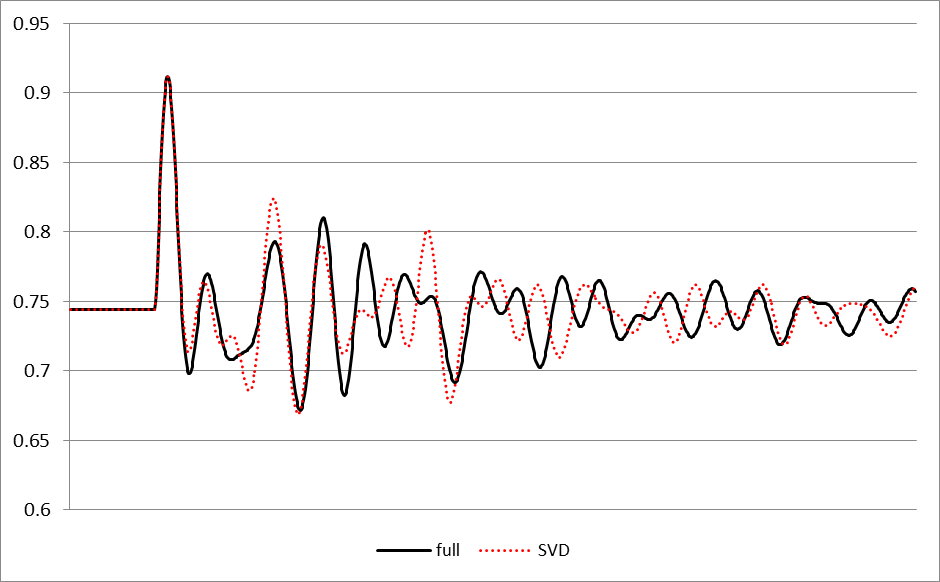}
      \caption{Full model vs. SVD}
    \end{subfigure}
    \hspace{0.05\textwidth}
    \begin{subfigure}[b]{0.45\textwidth}
      \includegraphics[width=\linewidth]{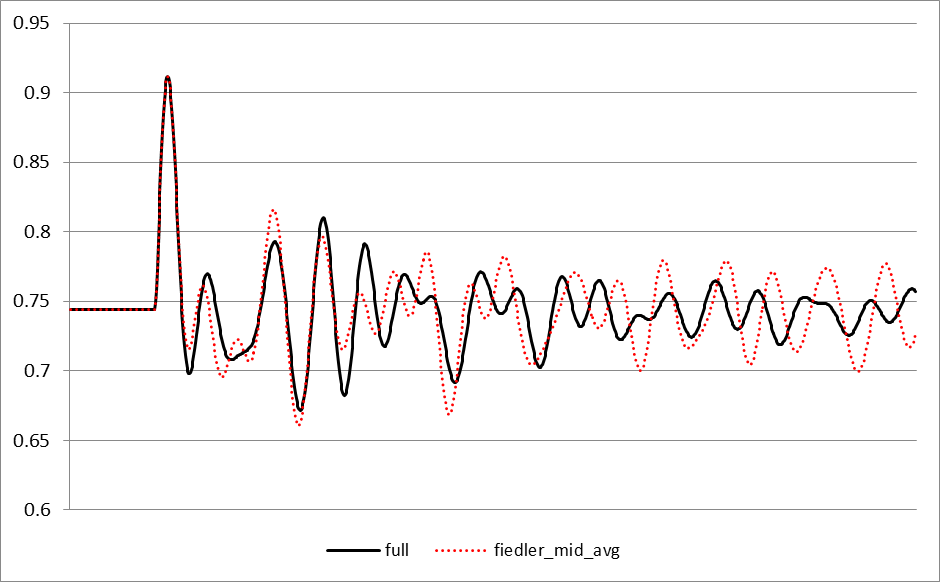}
      \caption{Full model vs. Fiedler-mid-avg}
    \end{subfigure}
    \\
    \begin{subfigure}[b]{0.45\textwidth}
      \includegraphics[width=\linewidth]{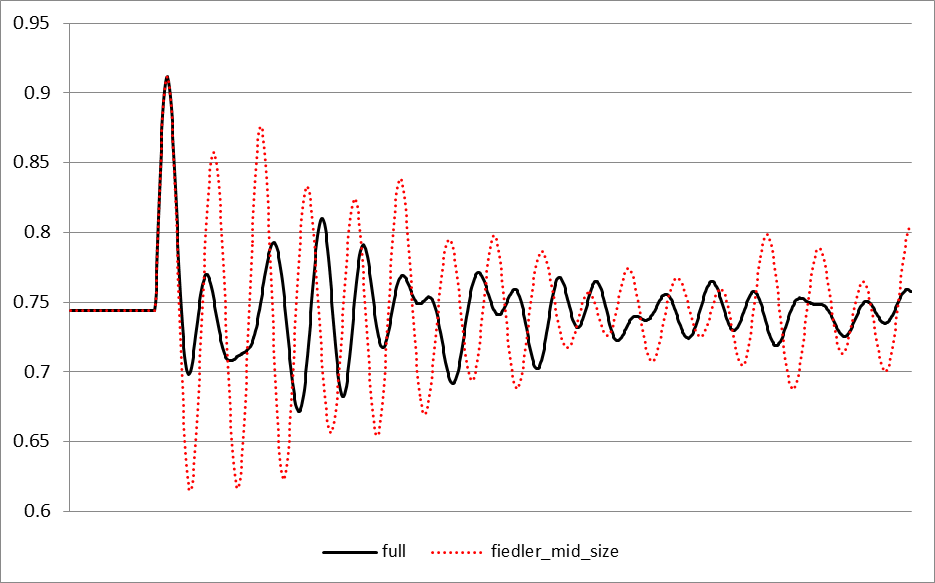}
      \caption{Full model vs. Fiedler-mid-size}
    \end{subfigure}
    \hspace{0.05\textwidth}
    \begin{subfigure}[b]{0.45\textwidth}
      \includegraphics[width=\linewidth]{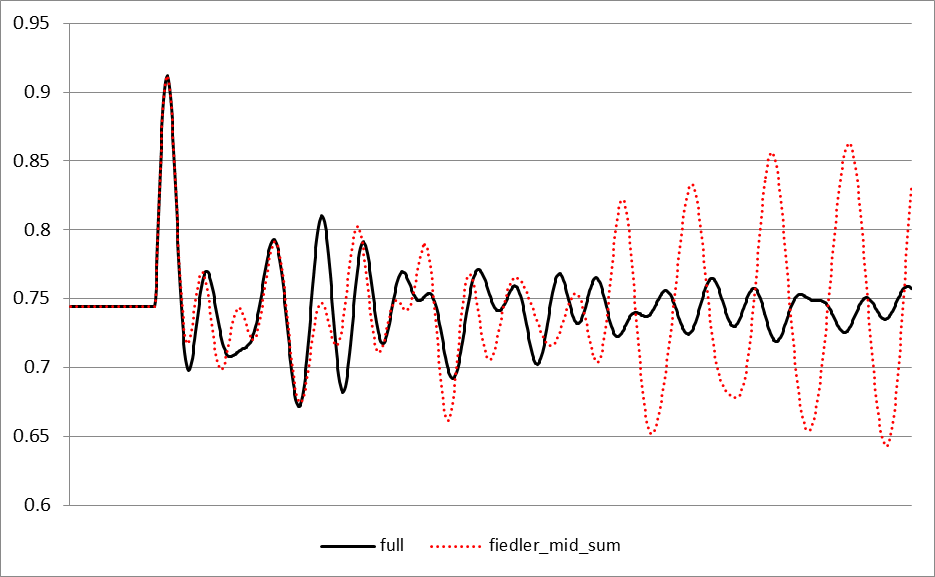}
      \caption{Full model vs. Fiedler-mid-sum}
    \end{subfigure}
    \caption{PMU traces for generator 29 in system $S_1$ with the first fault scenario under the full model
    (no reduction) and three separate reduced models for $r=10$.}\label{fig:pmu_29}
\end{figure}

Generator 45 is chosen because most methods seem to perform fairly well in
this fault scenario on it. In contrast, generator 29 is one where most
reductions perform fairly poorly. In Figure \ref{fig:pmu_29}, similar PMU
traces are shown for generator 29 in $S_1$ still under the first fault
scenario. It is much more difficult to judge a well-performing versus a
poorly-performing reduction just by looking at these particular PMU trace
plots, which is one of the reasons for choosing the more global perfprof
areas as the comparison method.

\subsubsection{System $S_2$}

System $S_2$ features a much smaller set with only nine generators in the
external area. There still are a few degenerate cases where a method did not
return a reduction for some $r$ values, but it is not as widespread as in
system $S_1$. In addition, methods Fiedler-mid-avg and Fiedler-mid-sum always
return identical reductions. For completeness, both methods are shown in the
tables although they have exactly the same values.

As in the previous section, Table \ref{tab:S2Ja} contains the areas under
perfprof plots summaries. In this system, the three Fiedler-zero methods
perform poorly just as in $S_1$. Yet, there are two major differences between
$S_1$ and $S_2$. Instead of SVD on top, $k$-means leads by a wide margin,
whereas in $S_1$ $k$-means is not a top performer. Additionally,
Fiedler-mid-size does very poorly in $S_2$ compared with $S_1$.

\begin{table}[h]
\begin{center}
{\footnotesize
\begin{tabular}{l|llll|l}
Method \textbackslash~ $r$ &	3&	4&	5&	6&	avg\\\hline
$k$-means	&0.816	&0.984	&0.988	&0.933	&0.931  \\
SVD	&0.714	&0.967	&0.938	&0.822	&0.860   \\
Fiedler-mid-avg	&0.718	&0.952	&0.798	&0.779	&0.812     \\
Fiedler-mid-sum	&0.718	&0.952	&0.798	&0.779	&0.812   \\
Spectral	&0.646	&0.962	&0.900	&0.479	&0.747        \\
Fiedler-zero-sum	&0.350	&0.943	&0.950	&0.684	&0.732 \\
Fiedler-zero-avg	&0.350	&0.943	&0.950	&0.684	&0.732  \\
Fiedler-zero-size	&0.636	&0.934	&0.501	&0.688	&0.689  \\
Fiedler-mid-size	&0.618	&0.640	&0.493	&0.656	&0.602
\end{tabular}
}
\end{center}
\caption{Relative areas under the perfprof curves for system $S_2$, ordered by average value. }\label{tab:S2Ja}
\end{table}
%

\begin{figure}[h]
\begin{center}
  \includegraphics[width=0.5\textwidth]{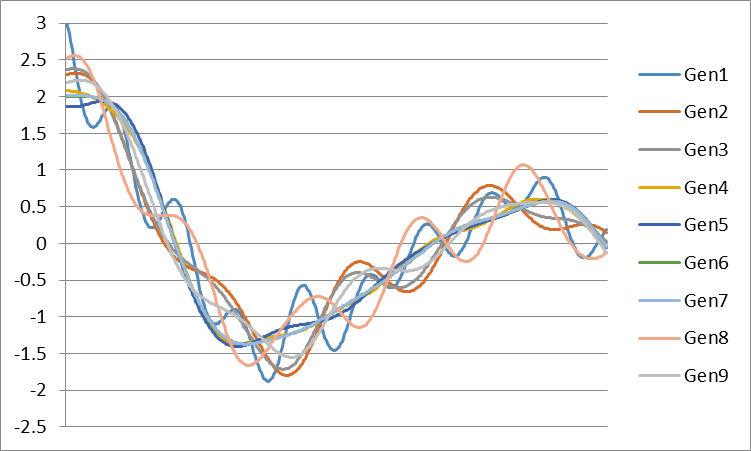}
  \caption{The PMU phasor values for all nine generators in the internal system
  for $S_2$ over 380 readings, totalling 3.8 seconds, following fault scenario 1.}
  \label{fig:S2F1_all}
\end{center}
\end{figure}

\begin{figure}[h]
\centering
    \begin{subfigure}[b]{0.45\textwidth}
      \includegraphics[width=\linewidth]{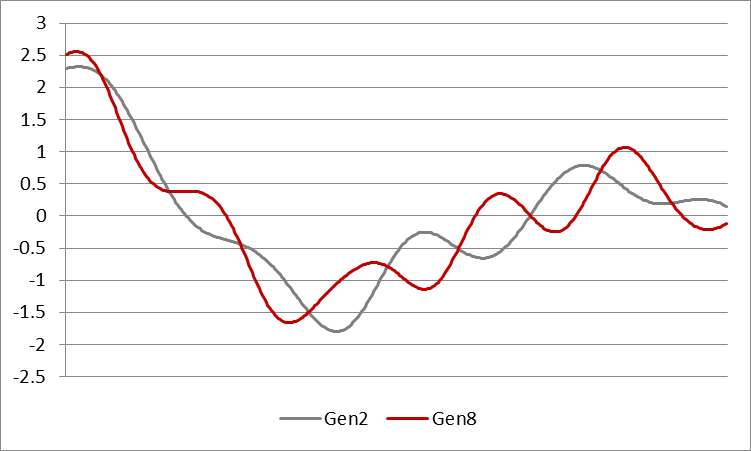}
      \caption{}
    \end{subfigure}
    \hspace{0.05\textwidth}
    \begin{subfigure}[b]{0.45\textwidth}
      \includegraphics[width=\linewidth]{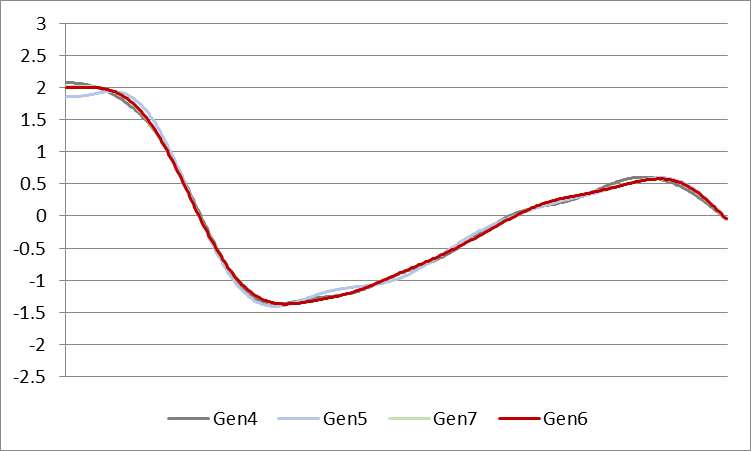}
      \caption{}
    \end{subfigure}
    \\
    \begin{subfigure}[b]{0.45\textwidth}
      \includegraphics[width=\linewidth]{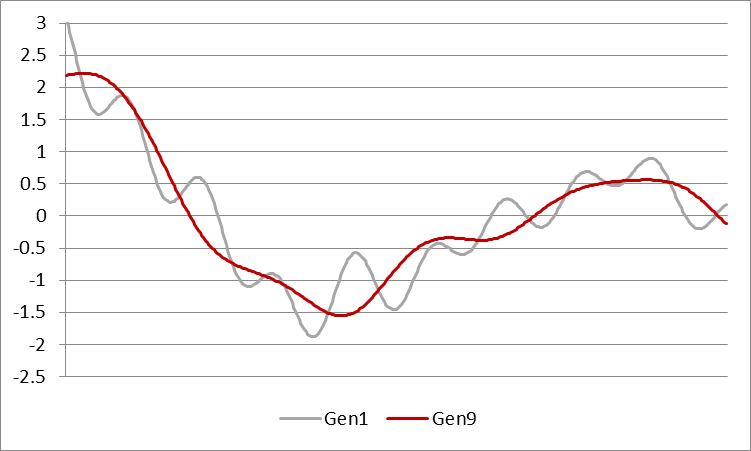}
      \caption{}
    \end{subfigure}
    \hspace{0.05\textwidth}
    \begin{subfigure}[b]{0.45\textwidth}
      \includegraphics[width=\linewidth]{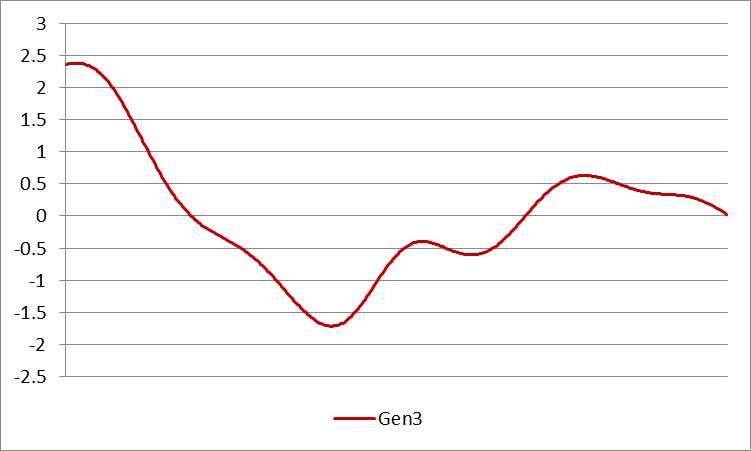}
      \caption{}
    \end{subfigure}
    \caption{Clustering via Fiedler-mid-sum for system $S_2$ following fault
    scenario 1 with $r=4$. Representative generators are in red in each plot.}
    \label{fig:S2F1_fmsu}
\end{figure}

\begin{figure}[h]
\centering
    \begin{subfigure}[b]{0.45\textwidth}
      \includegraphics[width=\linewidth]{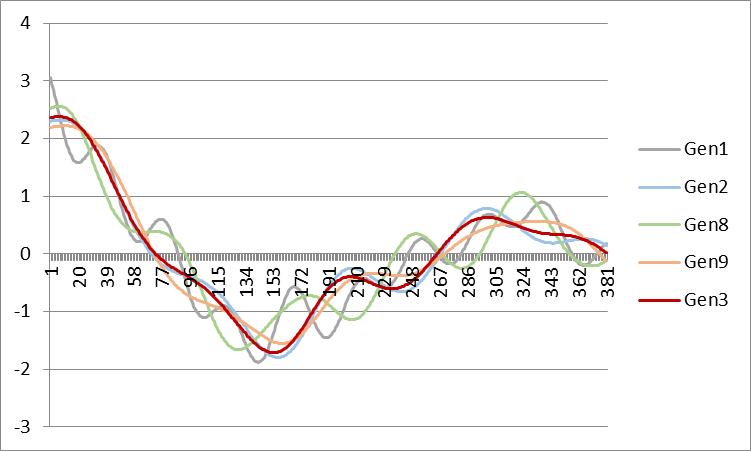}
      \caption{}
    \end{subfigure}
    \hspace{0.05\textwidth}
    \begin{subfigure}[b]{0.45\textwidth}
      \includegraphics[width=\linewidth]{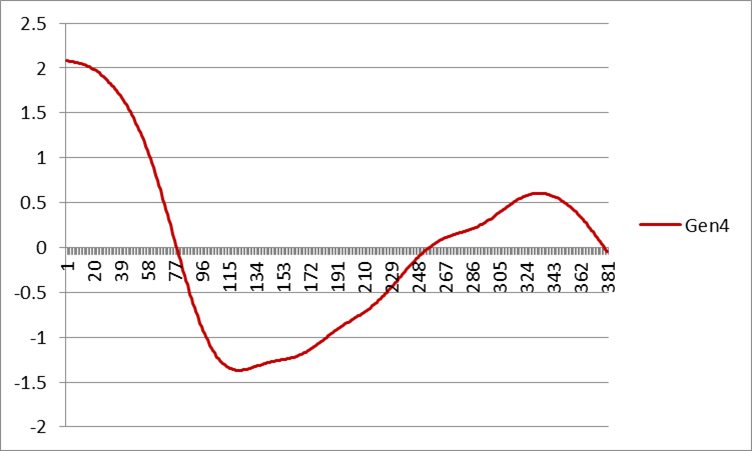}
      \caption{}
    \end{subfigure}
    \\
    \begin{subfigure}[b]{0.45\textwidth}
      \includegraphics[width=\linewidth]{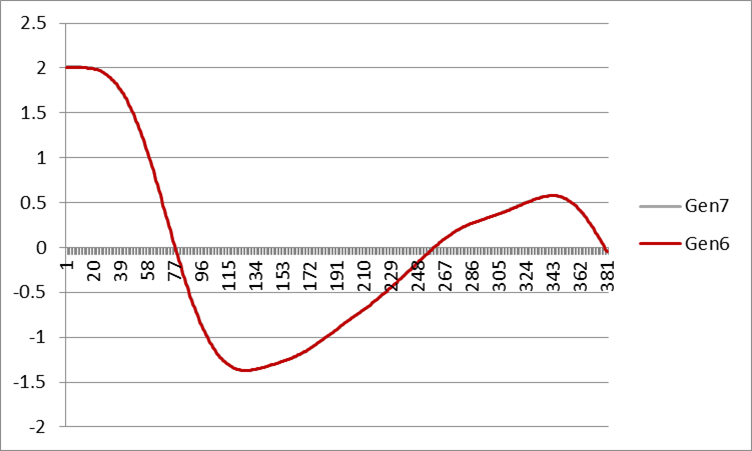}
      \caption{}
    \end{subfigure}
    \hspace{0.05\textwidth}
    \begin{subfigure}[b]{0.45\textwidth}
      \includegraphics[width=\linewidth]{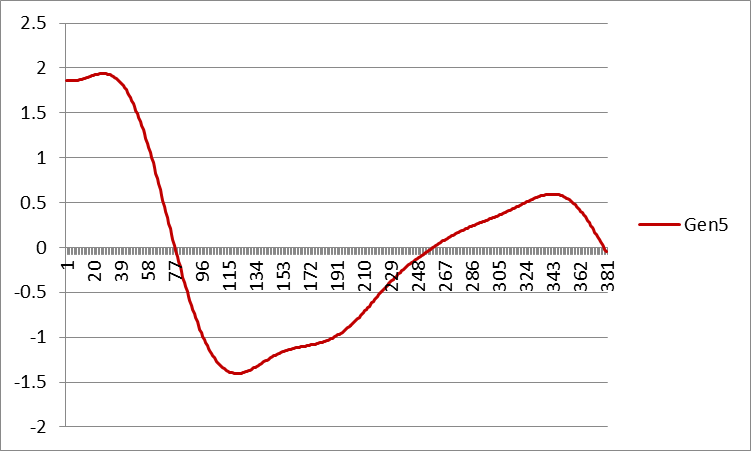}
      \caption{}
    \end{subfigure}
    \caption{Clustering via $k$-means for system $S_2$ following fault
    scenario 1 with $r=4$. Representative generators are in red in each plot.}
    \label{fig:S2F1_kmeans}
\end{figure}

Rather than providing PMU traces, an example partitioning of the input data
is shown. In Figure \ref{fig:S2F1_all}, the full set of PMU phasor values
recorded at all nine internal generators are displayed on the same axes.
Then, Figures \ref{fig:S2F1_fmsu} and \ref{fig:S2F1_kmeans} show the
clustering into four sets of generators for Fiedler-mid-sum and $k$-means,
respectively. Although $k$-means did perform better overall in $S_2$, looking
at the clusters themselves one could argue that the representative generators
found using the Fiedler clustering seem more representative of the system as
a whole. In particular, $k$-means split up generators 4 and 5 from 6 and 7
even though they appear to be very similar as shown in Figure
\ref{fig:S2F1_fmsu}(b). Thus, having generators 4, 5, and 6 as representative
generators in the $k$-means reduction might be redundant.

\section{Conclusion}\label{sec:conclusion}

In this paper, we provide a survey of three clustering techniques and an SVD
algorithm for dynamic model reduction along with a comparison of these
methods against two test systems: the IEEE 50 and IEEE 16 generator systems.
We compare two graph methods -- recursive spectral bipartitioning and
spectral clustering -- as well as an SVD method and the standard $k$-means
clustering algorithm. Our analysis (detailed in Section \ref{sec:results})
leads us to the following conclusions.

First, we remark that the $k$-means algorithm does not appear to be very
reliable. Although it does perform well on the smaller system but did not
rank highly in the larger system, we must conclude that $k$-means either is
not as well suited for larger systems, or it is not expected to consistently
work well. The latter statement is consistent with known problems using the
$k$-means algorithm: the data must be sufficiently separated and distributed
to yield useful conclusions. Thus, it seems that $k$-means, though sometimes
quite suited to this problem, may not always be dependable.

In contrast, we observe that the SVD method does seem to be a persistently
high performer. In $S_1$, SVD dominates almost all of the $r$ values and has
similarly high area under the perfprof curve in $S_2$. This consistency
across multiple systems will be necessary in a broadly applicable dynamic
model reduction algorithm. Similarly, the Fiedler-mid-avg and Fiedler-mid-sum
methods also appear to be repeatedly well-performing reduction techniques.

There also are three perpetually poorly performing algorithms: the
Fiedler-zero methods. Recall from Section \ref{sec:fiedler} that the standard
method of Fiedler partitioning is to split the set of vertices according to
positive or negative Fiedler value. However, in this application, it is clear
that splitting at the midpoint of the Fiedler values is much more
advantageous. Especially in the larger $S_1$ system, splitting at the zero
point can yield a degenerate partition. Even in the smaller $S_2$ system,
despite not creating degenerate partitions, the Fiedler-zero methods also
perform poorly.

Lastly, the spectral method ranks in the middle of our list of average area
under the curve for both test systems. However, looking at its performance on
specific $r$ values, there is high fluctuation. Recall that the final step of
the spectral clustering method involves using $k$-means. Given our earlier
conclusions that $k$-means may not be consistently reliable, it is not
unexpected to draw the same conclusion about the spectral clustering method.

Overall, given the analysis of these specific methods on these two relatively
small systems, we recommend that (a) dynamic model reduction techniques
should not rely on any kind of $k$-means clustering tools; (b) SVD is useful
if a quick solution is needed, there are many widely available software
packages for calculating SVDs, and it tended to be repeatedly
well-performing, if perhaps not always the top method; (c) if more time is
available for a test study on a specific system of interest, we suggest that
SVD be compared against Fiedler-mid-sum and Fiedler-mid-avg methods to see
which performs best on that system for a desired $r$ value. This paper
provides a useful method for comparing performance using performance profiles
given full and reduced model simulation data.


\section{Acknowledgements}
This work was supported in part by the Applied Mathematics Program of the
Office of Advanced Scientific Computing Research within the Office of Science
of the U.S. Department of Energy (DOE) through the Multifaceted Mathematics
for Complex Energy Systems (M2ACS) project.

The authors wish to thank the reviewers of this paper for their helpful
comments and suggestions which substantially improved the manuscript.

\vspace{-0.25cm}
\bibliographystyle{plain}
\bibliography{modRed}

\begin{thebibliography}{10}

\bibitem{Antoulas01asurvey}
A.~C. Antoulas, D.~C. Sorensen, and S.~Gugercin.
\newblock A survey of model reduction methods for large-scale systems.
\newblock {\em Contemp. Math.}, 280:193--219, 2001.

\bibitem{Bai20029}
Z.~Bai.
\newblock Krylov subspace techniques for reduced-order modeling of large-scale
  dynamical systems.
\newblock {\em Appl. Numer. Math.}, 43(1-–2):9--44, 2002.

\bibitem{genomicsSVD}
N.~M. Bertagnolli, J.~A. Drake, J.~M. Tennessen, and O.~Alter.
\newblock {SVD Identifies Transcript Length Distribution Functions from DNA
  Microarray Data and Reveals Evolutionary Forces Globally Affecting GBM
  Metabolism}.
\newblock {\em PLoS One}, 8(11):1--18, 2013.

\bibitem{canizares_linear_2004}
C.~A. Canizares, N.~Mithulananthan, F.~Milano, and J.~Reeve.
\newblock Linear performance indices to predict oscillatory stability problems
  in power systems.
\newblock {\em {IEEE} Trans. Power Syst.}, 19(2):1104--1114, 2004.

\bibitem{ChowAccariPrice}
J.~Chow, P.~Accari, and W.~Price.
\newblock Inertial and slow coherency aggregation algorithms for power system
  dynamic model reduction.
\newblock {\em IEEE Trans. Power Syst.}, 10(2):680–--685, 1995.

\bibitem{DiNHiN2013}
N.~J. Dingle and N.~J. Higham.
\newblock Reducing the influence of tiny normwise relative errors on
  performance profiles.
\newblock {\em ACM T. Math. Software}, 39, 2013.

\bibitem{DoEMoJ2002}
E.~D. Dolan and J.~J. Moore.
\newblock Benchmarking optimization software with performance profiles.
\newblock {\em Math. Program.}, 91:201--213, 2002.

\bibitem{fiedler1}
M.~Fiedler.
\newblock Algebraic connectivity of graphs.
\newblock {\em Czech Math. J.}, 23, 1973.

\bibitem{fiedler2}
M.~Fiedler.
\newblock A property of eigenvectors of nonnegative symmetric matrices and its
  application to graph theory.
\newblock {\em Czech Math. J.}, 25:619--633, 1975.

\bibitem{Freund1999}
R.~W. Freund.
\newblock {\em Reduced-Order Modeling Techniques Based on Krylov Subspaces and
  Their Use in Circuit Simulation}, pages 435--498.
\newblock Birkh{\"a}user Boston, Boston, MA, 1999.

\bibitem{GeJ1998}
J.~E. Gentle.
\newblock {\em Numerical Linear Algebra for Applications in Statistics},
  chapter Singular Value Factorization, $\S$ 3.2.7, pages 102--103.
\newblock Springer-Verlag, 1998.

\bibitem{GoGVaC1996}
G.~H. Golub and C.~F.~Van Loan.
\newblock {\em Matrix Computations, $3^{rd}$ edition}.
\newblock The Johns Hopkins University Press, Baltimore, {MD}, 1996.

\bibitem{HiDHiN2005}
D.~J. Higham and N.~J. Higham.
\newblock {\em {MATLAB Guide, Second Edition}}.
\newblock SIAM, 2005.

\bibitem{hogan_towards_2013}
E.~Hogan, E.~Cotilla-Sanchez, M.~Halappanavar, S.~Wang, P.~Mackey, P.~Hines,
  and Z.~Huang.
\newblock Towards effective clustering techniques for the analysis of electric
  power grids.
\newblock In {\em Proc. 3rd Workshop on HiPCNA-PG}, 2013.

\bibitem{honarkhah_stochastic_2010}
M.~Honarkhah and J.~Caers.
\newblock Stochastic simulation of patterns using distance-based pattern
  modeling.
\newblock {\em Math. Geosci.}, 42(5):487--517, 2010.

\bibitem{venkat_analysis_2004}
D.~N. Kosterev, C.~W. Taylor, and W.~A. Mittelstadt.
\newblock Model validation for the {A}ugust 10, 1996 {WSCC} system outage.
\newblock {\em IEEE Trans. on Power Syst.}, 14:967--979, 1999.

\bibitem{LiDWuX09}
D.~Lin and X.~Wu.
\newblock Phrase clustering for discriminative learning.
\newblock In {\em Annual Meeting of the ACL and IJCNLP}, pages 1030--1038,
  2009.

\bibitem{luxburg_tutorial_2007}
U.~von Luxburg.
\newblock A tutorial on spectral clustering.
\newblock {\em Stat. Comput.}, 17(4):395--416, 2007.

\bibitem{Pinnau2008}
R.~Pinnau.
\newblock {\em Model Reduction via Proper Orthogonal Decomposition}, pages
  95--109.
\newblock Springer Berlin Heidelberg, Berlin, Heidelberg, 2008.

\bibitem{Pothen:1990:PSM:84514.84521}
A.~Pothen, H.~D. Simon, and K-P Liou.
\newblock Partitioning sparse matrices with eigenvectors of graphs.
\newblock {\em SIAM J. Matrix Anal. Appl.}, 11(3):430--452, 1990.

\bibitem{rogers_power_2000}
G.~Rogers.
\newblock {\em Power System Oscillations}.
\newblock Power Electronics and Power Systems. Springer, 2000.

\bibitem{rudnick_power-system_1981}
H.~Rudnick, R.~I. Patino, and A.~Brameller.
\newblock Power-system dynamic equivalents: coherency recognition via the rate
  of change of kinetic energy.
\newblock {\em {IEEE} Proc.-C}, 128(6):325--333, 1981.

\bibitem{Sahidullah20161}
M.~Sahidullah and T.~Kinnunen.
\newblock Local spectral variability features for speaker verification.
\newblock {\em Digit. Signal Process.}, 50:1--11, 2016.

\bibitem{sastry_coherency_1981}
S.~Sastry and P.~Varaiya.
\newblock Coherency for interconnected power systems.
\newblock {\em {IEEE} Trans. Autom. Control}, {AC}-26(1):218--226, 1981.

\bibitem{schaeffer_graph_2007}
S.~E. Schaeffer.
\newblock Graph clustering.
\newblock {\em Computer Science Review}, 1(1):27--64, 2007.

\bibitem{selim_k-means-type_1984}
S.~Z. Selim and M.~A Ismail.
\newblock K-means-type algorithms: A generalized convergence theorem and
  characterization of local optimality.
\newblock {\em {IEEE} Trans. Pattern Anal. Mach. Intell}, {PAMI}-6(1):81--87,
  1984.

\bibitem{sun_online_2007}
K.~Sun, S.~Likhate, V.~Vittal, V.~S. Kolluri, and S.~Mandal.
\newblock An online dynamic security assessment scheme using phasor
  measurements and decision trees.
\newblock {\em {IEEE} Trans. Power Syst.}, 22(4):1935--1943, 2007.

\bibitem{tiako_real-time_2012}
R.~Tiako, D.~Jayaweera, and S.~Islam.
\newblock Real-time dynamic security assessment of power systems with large
  amount of wind power using case-based reasoning methodology.
\newblock In {\em 2012 {IEEE} Power and Energy Society General Meeting}, pages
  1--7, July 2012.

\bibitem{wang_measurement-based_2012}
S.~Wang, S.~Lu, G.~Lin, and N.~Zhou.
\newblock Measurement-based coherency identification and aggregation for power
  systems.
\newblock In {\em 2012 IEEE Power and Energy Society General Meeting}, 2012.

\bibitem{wang_dynamic-feature_2014}
S.~Wang, S.~Lu, N.~Zhou, G.~Lin, M.~Elizondo, and M.~A. Pai.
\newblock Dynamic-feature extraction, attribution, and reconstruction ({DEAR})
  method for power system model reduction.
\newblock {\em {IEEE} Trans. Power Syst.}, 29(5):2049--2059, 2014.

\bibitem{xue_new_1998}
Y.~Xue, Y.~Yu, J.~Li, Z.~Gao, C.~Ding, F.~Xue, L.~Wang, G.~K. Morison, and
  P.~Kundur.
\newblock A new tool for dynamic security assessment of power systems.
\newblock {\em Control Eng. Pract.}, 6(12):1511--1516, 1998.

\end{thebibliography}

\end{document}